\documentclass[a4paper,aps,prd,10pt,preprintnumbers,showpacs,twocolumn,superscriptaddress,nofootinbib,amsmath,amssymb]{revtex4-1}
\usepackage{graphicx}
\usepackage{cmap}
\usepackage[utf8]{inputenc}
\usepackage[T1]{fontenc}

\def\imo{i}

\def\K{{\cal K}}
\def\Order#1{{\cal O}\left(#1\right)}
\DeclareMathAlphabet{\pazocal}{OMS}{zplm}{m}{n}
\def\dfrac#1#2{\frac{\displaystyle #1}{\displaystyle #2}}

\begin{document}

\title{Analytic expressions for grey-body factors of the general parametrized spherically symmetric black holes}
\author{Alexey Dubinsky}\email{dubinsky@ukr.net}
\affiliation{Pablo de Olavide University, Seville, Spain}
\author{Antonina F. Zinhailo}\email{F170631@fpf.slu.cz}
\affiliation{Research Centre for Theoretical Physics and Astrophysics, \\ Institute of Physics, Silesian University in Opava, \\ Bezručovo náměstí 13, CZ-74601 Opava, Czech Republic}

\begin{abstract}
In light of the recently discovered connection between grey-body factors and quasinormal ringing, we derive analytic expressions for the grey-body factors of generic parametrized spherically symmetric and asymptotically flat black holes. These expressions are presented as expansions in terms of the inverse multipole number and the coefficients of the parametrization. The obtained analytic formulas serve as good approximations whenever the deviation from the Schwarzschild geometry is not very large. We demonstrate that the primary parameter determining the grey-body factors is the deviation of the event horizon radius from its Schwarzschild value, while the higher-order coefficients of the parametrization, which govern the near-horizon geometry, are much less significant. This finding is consistent with recent observations that grey-body factors are considerably more stable against small deformations of the near-horizon geometry than quasinormal modes.
\end{abstract}

\maketitle

\textbf{Introduction.} Grey-body factors of black holes are quantities that describe how the black hole modifies the spectrum of radiation passing through its gravitational field before escaping to infinity. They play a crucial role in the study of black hole thermodynamics \cite{Bekenstein:1973ur} and Hawking radiation \cite{Hawking:1975vcx}. Unlike the idealized black body radiation predicted by Hawking, which assumes a perfect black body, real black holes partially absorb and reflect radiation. Grey-body factors quantify this deviation from perfect black body behavior.

Recently, it has been discovered a new aspect of the grey-body factors. It was shown that the black hole grey-body factor, $\Gamma_{\ell m} (\omega)$, plays a crucial role in estimating the remnant parameters from gravitational-wave quasinormal ringing \cite{Oshita:2023cjz}. Specifically, for $(\ell, m) = (2, 2)$, where $\ell$ and $m$ are the multipole and azimuthal quantum numbers, respectively, there is a connection between the grey-body factors $\Gamma_{\ell m}$ and the gravitational-wave spectral amplitude $|\tilde{h}_{\ell m} (\omega)|$ for $\omega \gtrsim f_{\ell m} \equiv \operatorname{Re}(\omega_{\ell m0})$, which can be expressed as follows:
\begin{equation}
|\tilde{h}_{\ell m} (\omega)| \simeq c_{\ell m} \times \gamma_{\ell m} (\omega) \equiv c_{\ell m} \times \omega^{-3}. \sqrt{1-\Gamma_{\ell m} (\omega)}.
\label{main}
\end{equation}
Here, $\omega_{\ell m 0}$ represents the frequency of the gravitational wave with the multipole and azimuthal numbers $\ell$ and $m$, implying the fundamental frequency $n=0$, and $c_{\ell m}$ is the gravitational-wave amplitude. This suggests that there may be an intrinsic link between the quasinormal modes of black holes \cite{Konoplya:2011qq, Kokkotas:1999bd} and their grey-body factors.

Moreover, it has been found that, unlike the first few overtones of the quasinormal mode spectrum \cite{Konoplya:2022pbc}, grey-body factors are significantly more stable against small deviations in the geometry. They deviate only slightly from the Kerr limit when the geometry is slightly deformed from the Kerr configuration \cite{Rosato:2024arw, Oshita:2024fzf}. A general correspondence between the grey-body factors of spherically symmetric and asymptotically flat or de Sitter black holes and their quasinormal modes has been developed in \cite{Konoplya:2024lir} and extended to axially symmetric black hole \cite{Konoplya:2024vuj}. Additionally, a specific link between these two quantities for the Kerr spacetime has been discussed in \cite{Oshita:2024wgt}.

While sufficiently accurate calculations of grey-body factors primarily rely on numerical methods \cite{Page:1976df, Page:1976ki, Page:1977um, Kanti:2004nr, Grain:2005my, Kanti:2014dxa, Pappas:2016ovo}, in this work, we derive analytic expressions for grey-body factors that are sufficiently accurate over a wide range of parameters. Instead of focusing on specific black hole metrics, we use a general parametrized form of spherically symmetric and asymptotically flat black holes developed in \cite{Rezzolla:2014mua}. We assume that deviations from the Schwarzschild geometry are not significant, allowing the parameters of the parametrization (which are zero in the Schwarzschild limit) to be considered small. Previous works on quasinormal modes and optical phenomena around such general parametrized black holes \cite{Konoplya:2020hyk, Konoplya:2021slg} indicate that the infinite series in the parametrization can usually be truncated after the first few terms without loss of accuracy.

To find grey-body factors, we use the WKB method, which has been applied in \cite{Konoplya:2010kv, Kokkotas:2010zd}, demonstrating reasonable agreement with accurate results whenever the multipole number is greater than zero. To test the accuracy of the obtained analytic formulas, we use numerical integration of the wave equation to obtain precise grey-body factors. For deriving analytic expressions of the grey-body factors, we expand the WKB formula in terms of inverse multipole numbers beyond the eikonal limit as prescribed in \cite{Konoplya:2023moy}. This method has been recently applied to find analytic expressions for quasinormal modes in some particular theories of gravity \cite{Malik:2024sxv, Malik:2024voy, Bolokhov:2024ixe, Malik:2024tuf,2753764,Malik:2024nhy}.

\textbf{Parametrized black hole metric.} The spherically symmetric black hole can be described by the following general metric, which incorporates two independent metric functions $f(r)$ and $g(r)$,
\begin{equation}
ds^2=-f(r)dt^2+g^{-1}(r)dr^2+r^2 (d\theta^2+\sin^2\theta d\phi^2)\label{metric}.
\end{equation}
The usual way is to use instead the functions $B(r)$ and $N(r)$, such that $f(r)= N^2(r)$, $g(r) = N^2(r)/B^2(r)$. Here $r_0$ is the radius of the event horizon: $N(r_0)=0$.

The general parameterization for spherically symmetric and asymptotically flat black holes was proposed in \cite{Rezzolla:2014mua}, and extended to axial symmetry in \cite{Konoplya:2016jvv}. It has been further applied and discussed in numerous works \cite{Kocherlakota:2020kyu,Zhang:2024rvk,Cassing:2023bpt,Li:2021mnx,Ma:2024kbu,Bronnikov:2021liv,Shashank:2021giy,Konoplya:2021slg,Kokkotas:2017zwt,Yu:2021xen,Toshmatov:2023anz,Nampalliwar:2019iti,Ni:2016uik,Konoplya:2018arm,Zinhailo:2018ska,Paul:2023eep}, so here we will only briefly review its key features.

Following \cite{Rezzolla:2014mua}, we employ the dimensionless compact variable
$x \equiv 1 - (r_0/r)$.
In terms of this compact coordinate, $x=0$ corresponds to the event horizon, and $x=1$ corresponds to infinity. It is standard practice to rewrite the metric function $N$ as $N^2=x A(x)$, where $A(x)>0$ for \mbox{$0\leq x\leq1$}. The functions $A$ and $B$ are:
\begin{eqnarray}\nonumber
A(x)&=&1-\epsilon (1-x)+(a_0-\epsilon)(1-x)^2+{\tilde A}(x)(1-x)^3\,,
\\
B(x)&=&1+b_0(1-x)+{\tilde B}(x)(1-x)^2\,.\label{ABexp}
\end{eqnarray}
Here, the coefficient $\epsilon$ measures the deviation of the event horizon radius $r_0$ from its Schwarzschild limit $2 M$, i. e.
$\epsilon = (2 M-r_0)/r_0.$
The coefficients $a_0$ and $b_0$ can be expressed in terms of the post-Newtonian (PN) parameters,
$$
a_0=(\beta-\gamma)(1+\epsilon)^2/2,
\quad
b_0=(\gamma-1)(1+\epsilon)/2.
$$
Current observational constraints imply \mbox{$a_0 \sim b_0 \sim 10^{-4}$},  so that these coefficients can be neglected.

The functions ${\tilde A}$ and ${\tilde B}$ define the geometry in the near horizon region (i.e., for $x \simeq 0$) and are represented by an infinite continued fraction as follows:
\begin{equation}\label{ABdef}
{\tilde A}(x)=\dfrac{a_1}{1+\dfrac{a_2x}{1+\dfrac{a_3x}{1+\ldots}}}, \quad
{\tilde B}(x)=\dfrac{b_1}{1+\dfrac{b_2x}{1+\dfrac{b_3x}{1+\ldots}}}.
\end{equation}
Here $a_1, a_2,\ldots$ and $b_1, b_2,\ldots$ are dimensionless constants - the coefficients of the parameterization. Assuming $a_{0}=b_{0}=0$, at the event horizon only the first term in each of the continued fractions remains:
$ 
{\tilde A}(0)={a_1},~
{\tilde B}(0)={b_1}.
$ 
This relationship implies that near the event horizon only the lower-order terms of the expansions are significant, while the higher-order terms can generally be ignored. Therefore, the above infinite continued fraction can be truncated to include only the first few parameters, which already serves as a good approximation  \cite{Konoplya:2020hyk}.

\textbf{Wave equation and effective potentials.} Minimally coupled scalar (given by the scalar function $\Phi$) and electromagnetic (given by the four-vector potential $A_\mu$) fields obey the well-known general covariant Klein-Gordon and Maxwell equations.
Using the expansion into the standard spherical harmonics, the perturbation equations can be reduced to  the universal radial wavelike form \cite{Abdalla:2006qj}:
\begin{equation}\label{wave-equation}
\dfrac{d^2 \Psi}{dr_*^2}+(\omega^2-V(r))\Psi=0,
\end{equation}
where $r_*$ is ``tortoise coordinate''
$dr_*\equiv  (f(r) g(r))^{-1/2} d r$.
The corresponding effective potentials for scalar ($s=0$) and electromagnetic ($s=1$) fields are
\begin{equation}\label{potentialScalar}
V(r)=f(r)\frac{\ell(\ell+1)}{r^2}+\frac{1-s}{r}\cdot\frac{d^2 r}{dr_*^2},
\end{equation}
where $\ell$ is multipole number, keeping also the generation of the azimuthal quantum number $m$.

\begin{figure}
\resizebox{0.8\linewidth}{!}{\includegraphics{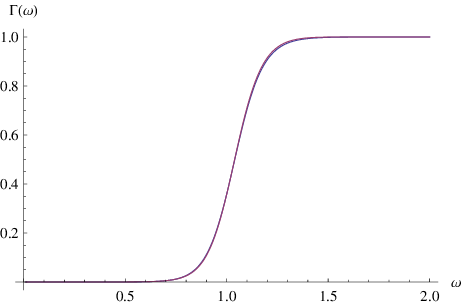}}\\
\resizebox{0.8\linewidth}{!}{\includegraphics{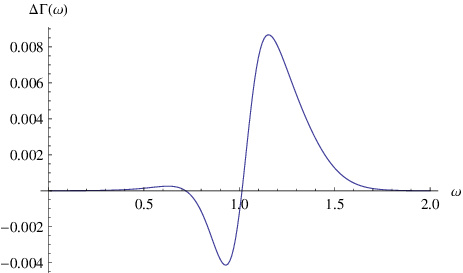}}
\caption{Transmission coefficients for scalar field perturbations with $\epsilon=-0.1$, $a_1 = 0.1$, $b_1 = 0.1$ at $\ell=2$, calculated using both the analytic formula and the 6th-order WKB approximation (up), along with the difference between the two methods (down).}\label{fig:L2s0}
\end{figure}
For a general parametrized black hole that is not associated with any specific theory, it is impossible to write down the wave equations for gravitational perturbations in a universal form. However, various theories use an effective approach based on the assumption that the dynamics of the gravitational field can be described by Einstein's theory of gravity with an effective energy-momentum tensor representing an anisotropic fluid \cite{Ashtekar:2018lag,Ashtekar:2018cay}. 
After all, there are Einsteinian theories with anisotropic fluid producing black hole solutions \cite{Dehnen:2003rc,Ivashchuk:2010sn}. 
In all such scenarios, axial gravitational perturbations ($s=2$) can be reduced to the following form (see, for example, \cite{Bouhmadi-Lopez:2020oia,Konoplya:2024lch,Dubinsky:2024rvf}).
\begin{equation}
V(r)= f(r)\left(\frac{2g(r)}{r^2}-\frac{(fg)'}{2rf}+\frac{(\ell+2)(\ell-1)}{r^2}\right).
\end{equation}
All three effective potentials for scalar, electromagnetic and effective gravitational perturbations can be cast to a universal form
\begin{align}\nonumber
V_{s}(r) = & \frac{(1-s) \left(g(r) f'(r) + f(r) g'(r)\right)}{2r} - \\
& \frac{(1-s) s (f(r) g(r) - f(r))}{r^2}  + \frac{\ell (\ell + 1) f(r)}{r^2}.
\end{align}

\textbf{WKB approach.} The WKB method is popular and efficient method for finding quasinormal modes and grey-body factors which was used in numerous publications (see, for instance, \cite{Matyjasek:2024uwo,Matyjasek:2021xfg,Konoplya:2005sy,Skvortsova:2024wly,Dubinsky:2024aeu,Dubinsky:2024fvi,Konoplya:2001ji,Konoplya:2006ar,Bolokhov:2023bwm,Kodama:2009bf}). It relies on matching the WKB asymptotic solutions with the Taylor expansion around the peak of the potential barrier. The first-order WKB formula corresponds to the eikonal approximation and becomes exact in the limit $\ell \rightarrow \infty$. The general WKB expression for the quasinormal frequencies can then be written as an expansion around the eikonal limit as follows \cite{Konoplya:2019hlu}:
\begin{eqnarray}\label{WKBformula-spherical}
\omega^2&=&V_0+A_2(\K^2)+A_4(\K^2)+A_6(\K^2)+\ldots\\\nonumber
&-&\imo\K\sqrt{-2V_2}\left(1+A_3(\K^2)+A_5(\K^2)+A_7(\K^2)\ldots\right),
\end{eqnarray}
The explicit forms of $A_i$ care written out explicitly in \cite{Iyer:1986np} for the second and third WKB orders, in \cite{Konoplya:2003ii} for the 4th-6th orders, and in \cite{Matyjasek:2017psv} for the 7th-13th orders. The same WKB corrections are used for the scattering boundary problem, but with a different resultant expression for $\K$.

In the scattering problem,  partial reflection by the potential barrier of the wave going from the horizon leads to the same grey-body factors as partial reflection of the wave going from infinity to the potential peak. Therefore, the boundary conditions have the form:
\begin{equation}
\begin{array}{rclcl}
\Psi &=& e^{-i\omega r_*} + R e^{i\omega r_*}, &\quad& r_*\to+\infty, \\
\Psi &=& T e^{-i\omega r_*}, &\quad& r_*\to-\infty,
\end{array}
\end{equation}
where $R$ and $T$ are the reflection and transmission coefficients respectively. Following \cite{Iyer:1986np},  $R$ and $T$ can be found as follows:
\begin{eqnarray}\label{reflection}
|R|^2&=&(1+e^{-2\pi\imo \K})^{-1}.\\
\label{transmission}
|T|^2&=&1-|R|^2=(1+e^{2\pi\imo \K})^{-1}.
\end{eqnarray}
Here $\K$ is a function of $\omega$, defined through  the WKB corrections of the same form as for quasinormal modes in (\ref{WKBformula-spherical}).

While analytic expressions, primarily for quasinormal modes rather than grey-body factors, have been derived in the eikonal limit—which has several intriguing aspects, such as the null geodesics/eikonal quasinormal correspondence \cite{Cardoso:2008bp,Konoplya:2017wot,Konoplya:2022gjp,Bolokhov:2023dxq} and eikonal instability \cite{Gleiser:2005ra,Takahashi:2011du,Konoplya:2017lhs}—the literature on expansions beyond the eikonal limit is sparse. It includes only a few works focused on quasinormal modes for specific black hole models \cite{Malik:2024voy, Bolokhov:2024ixe, Malik:2024tuf,2753764,Malik:2024nhy,Malik:2024sxv,Ma:2024kbu,Davey:2023fin}. To derive analytic expressions for grey-body factors explicitly in terms of the parametrization coefficients, we will employ an expansion in terms of $\kappa^{-1}$ beyond the eikonal limit. Leveraging the approach introduced in \cite{Konoplya:2023moy}, we perform an expansion of the effective potential in the following form:
\begin{equation}\label{potential-multipole}
V(r_*)=\kappa^2\left(H_{0}(r_*)+H_{1}(r_*) \kappa^{-1} +\dots\right).
\end{equation}
In this context, we define $\kappa$ as $\kappa \equiv \ell + \frac{1}{2}$, where $\ell$ represents the multipole number, taking values $\ell = s, s+1, s+2, \dots$. Following the procedure outlined in \cite{Konoplya:2023moy}, we further expand the series in powers of $\kappa^{-1}$.

As demonstrated in \cite{Konoplya:2023moy}, the first-order WKB approximation, given by equation (\ref{WKBformula-spherical}), yields an eikonal limit expression for $\K$ as:
$-\imo\K_0=(\omega^2-V_0)(-2V_2)^{-1/2},$
which is valid under the condition that the classical turning points are in close proximity, i.e.
$\delta\equiv\omega^2-\Omega\kappa^2\approx0$.
We are specifically interested in the regime where $\omega$ satisfies this condition, i.e., when $\K \approx \K_0 \approx 0$, as it ensures that the transmission coefficients can be meaningfully distinguished from their trivial limits of zero or unity. By expanding $\K$ around its eikonal approximation $\K_0$, it was shown in \cite{Konoplya:2023moy} that, under the assumption $\K=\K_0+\Order{\kappa^{-1}}$, one obtains the relation:
\[
\omega^2\kappa^{-2}-\Omega^2\equiv\delta \kappa^{-2}\approx-\imo\K_0\sqrt{-2V_2}\kappa^{-2}=\Order{\kappa^{-1}}.
\]
This allows, through the expansion of $\omega$, the determination of a corresponding expansion for $\K$.

\textbf{Grey-body factors.} Using the above procedure and further expansion in terms of small values of the coefficients of the parametrization $\epsilon$, $a_1$, $b_1$, etc. we can find analytic expressions for $\K$, and threreby, for the grey-body factors as a series in powers of $\kappa^{-1}$, $\epsilon$, $a_1$, $b_1$:
\begin{widetext}
\begin{align*}
\K &= \frac{i(79180457472 \kappa^2 \left( \frac{4A\kappa^2}{6561} + \omega^2 \right)(9C^4 - 16\kappa^2 C^2 + 636417 E \left( \frac{4A\kappa^2}{6561} + \omega^2 \right)F))}{10727424 A^5 \kappa^3}+O(\epsilon^{2}, a_{1}^{2}, b_{1}^{2}, \kappa^{-3}).
\end{align*}
Here we introduced supplementary quantities 
\begin{align*}
 A &= 9(-450\epsilon + 4 b_{1}(64\epsilon - 27) + 243) + 4 a_{1}(-645\epsilon + 4 b_{1}(97\epsilon - 45) + 243)\\
 B &= 4(8a_{1}(7\epsilon - 9) + 27(10\epsilon - 9))\kappa^2 + 6561\omega^2 \\
 C &= 223488(22\epsilon - 27)s^2 + 5363712\epsilon s + \frac{47436B^2}{A\kappa^2} + \frac{151337970\epsilon B^2}{A^2\kappa^2} \\
\end{align*}
\begin{align*}
&\quad - \frac{89104941 B^2}{A^2\kappa^2} - 1772384\epsilon + 192b_{1}(1552(20\epsilon - 9)s^2 - 3104(20\epsilon - 9)s \\
&\quad - \frac{286380\epsilon B^2}{A^2\kappa^2} + \frac{163539 B^2}{A^2\kappa^2} + 6208\epsilon - 5044) + 1215216 \\
D &= 32b_{1}(55872(17\epsilon - 9)s^2 - 111744(17\epsilon - 9)s + \frac{5662143 B^2}{A^2\kappa^2} - 505564\epsilon - 338724) \\
&\quad + 27(-74496(2\epsilon + 9)s^2 + 2011392\epsilon s + \frac{26259687\epsilon B^2}{A^2\kappa^2} - \frac{14391189 B^2}{A^2\kappa^2} \\
&\quad + 547856\epsilon + 218832) \\
E &= 2048(-645\epsilon + 4b_{1}(97\epsilon - 45) + 243)^2(8b_{1}(144(17\epsilon - 9)s^2 - 288(17\epsilon - 9)s - 1303\epsilon - 873) \\
&\quad - 27(48(2\epsilon + 9)s^2 - 1296\epsilon s - 353\epsilon - 141))a_{1}^3 \\
&\quad + 2304(-645\epsilon + 4b_{1}(97\epsilon - 45) + 243)(64(36(14524\epsilon^2 - 13599\epsilon + 3159)s^2 \\
&\quad - 72(14524\epsilon^2 - 13599\epsilon + 3159)s - 124880\epsilon^2 - 94869\epsilon + 62937)b_{1}^2 \\
&\quad - 36(288(4739\epsilon^2 - 2565\epsilon + 243)s^2 - 864(5240\epsilon^2 - 4275\epsilon + 891)s \\
&\quad - 5(111538\epsilon^2 + 67257\epsilon - 61479))b_{1} \\
&\quad - 27(144(3530\epsilon^2 - 12339\epsilon + 5103)s^2 + 1728\epsilon(1780\epsilon - 891)s + 389870\epsilon^2 + 171531\epsilon - 200475))a_{1}^2 \\
&\quad + 648(2048(64\epsilon - 27)(72(7996\epsilon^2 - 7389\epsilon + 1701)s^2 - 144(7996\epsilon^2 - 7389\epsilon + 1701)s + 416\epsilon^2 - 127665\epsilon + 55161)b_{1}^3 \\
&\quad - 41472(132224\epsilon^3 - 1796537\epsilon^2 + 1592784\epsilon + 72s^2(72824\epsilon^3 - 85852\epsilon^2 + 33615\epsilon - 4374) \\
&\quad - 8s(1827184\epsilon^3 - 2407716\epsilon^2 + 1063611\epsilon - 157464) - 363042)b_{1}^2 \\
&\quad + 216(1152(315580\epsilon^3 + 171396\epsilon^2 - 359397\epsilon + 98415)s^2 - 6912(499640\epsilon^3 - 616140\epsilon^2 + 248832\epsilon - 32805)s \\
&\quad + \frac{(3953\epsilon - 873)B^2}{9\kappa^2} + 16(5514110\epsilon^3 - 32035095\epsilon^2 + 27970272\epsilon - 6652854))b_{1} \\
&\quad + 243(2304(221500\epsilon^3 - 558360\epsilon^2 + 382725\epsilon - 78732)s^2 + 6912\epsilon(153500\epsilon^2 - 151740\epsilon + 37179)s \\
&\quad - \frac{(833\epsilon - 123)B^2}{3\kappa^2} - 32(2167000\epsilon^3 - 7133130\epsilon^2 + 5662143\epsilon - 1318761))a_{1} \
\end{align*}
\begin{align*}
F &= 12288(27 - 64\epsilon)^2(4(20\epsilon - 9)s^2 - 8(20\epsilon - 9)s + 16\epsilon - 13)b_{1}^3 \\
&\quad - 256(64\epsilon - 27)(144(4592\epsilon^2 - 3618\epsilon + 729)s^2 - 5184(376\epsilon^2 - 348\epsilon + 81)s + 245888\epsilon^2 - 314658\epsilon + 96957)b_{1}^2 \\
&\quad + 144(576(9200\epsilon^3 + 78732\epsilon^2 - 81648\epsilon + 19683)s^2 - 31104(8400\epsilon^3 - 11236\epsilon^2 + 4968\epsilon - 729)s \\
&\quad + \frac{(48\epsilon - 13)B^2}{\kappa^2} + 8(5814400\epsilon^3 - 10108476\epsilon^2 + 5767848\epsilon - 1082565))b_{1} \\
&\quad + 81\left(55296s\epsilon(27 - 50\epsilon)^2 + 2304s^2(22\epsilon - 27)(27 - 50\epsilon)^2 - 16(1142\epsilon - 783)(27 - 50\epsilon)^2 - \frac{(1730\epsilon - 621)B^2}{27\kappa^2}\right).
\end{align*}
\end{widetext}

\begin{figure}
\resizebox{0.8\linewidth}{!}{\includegraphics{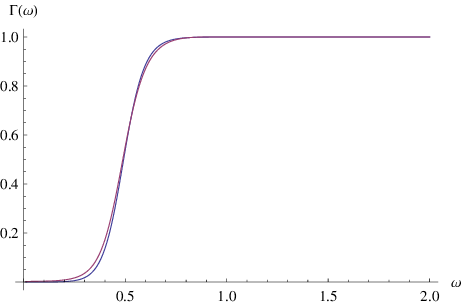}}\\
\resizebox{0.8\linewidth}{!}{\includegraphics{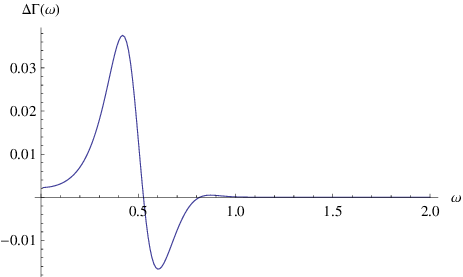}}
\caption{Transmission coefficients for electromagnetic field perturbations with $\epsilon=0.1$, $a_1 = 0.1$, $b_1 = 0.1$ at $\ell=1$, calculated using both the analytic formula and the 6th-order WKB approximation (up), along with the difference between the two methods (down).}\label{fig:L1s1}
\end{figure}
\begin{figure}
\resizebox{0.8\linewidth}{!}{\includegraphics{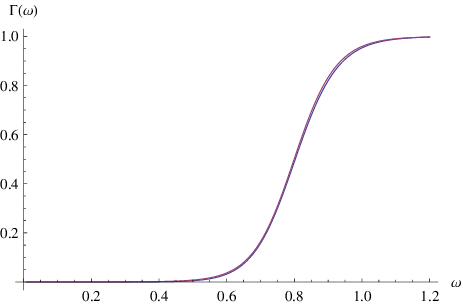}}\\
\resizebox{0.8\linewidth}{!}{\includegraphics{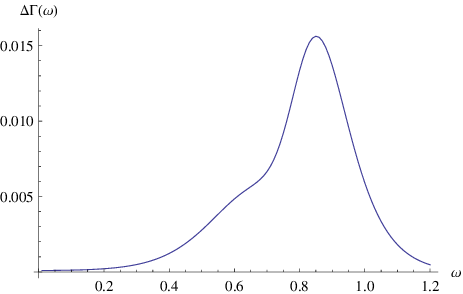}}
\caption{Transmission coefficients for effective axial gravitational perturbations with $\epsilon=-0.1$, $a_1 = 0.1$, $b_1 = 0.1$ at $\ell=2$, calculated using both the analytic formula expanded until the third order of $\kappa^{-1}$ and the 6th-order WKB approximation (up), along with the difference between the two methods (down).}\label{fig:L2s2}
\end{figure}
\begin{figure}
\resizebox{0.8\linewidth}{!}{\includegraphics{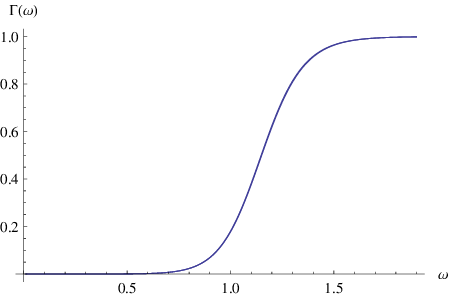}}\\
\resizebox{0.8\linewidth}{!}{\includegraphics{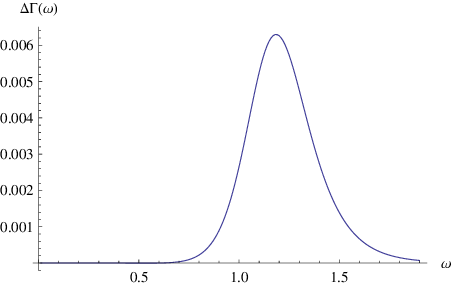}}
\caption{Up: Transmission coefficients for $\ell=1$ electromagnetic perturbations with $\epsilon=-0.5$, $a_1 = 0.1$, $b_1 = 0.1$ and two different values of $a_2$: $a_{2}=0$ and $a_{2}=1$ calculated with the 6th order WKB method. Down: the difference between grey-body factors of the two cases $a_{2}=0$ and $a_{2}=1$. Even relatively  large $a_2=1$ induce relatively small change in the grey-body factor.}\label{fig:a2}
\end{figure}
\begin{figure*}
\resizebox{\linewidth}{!}{\includegraphics{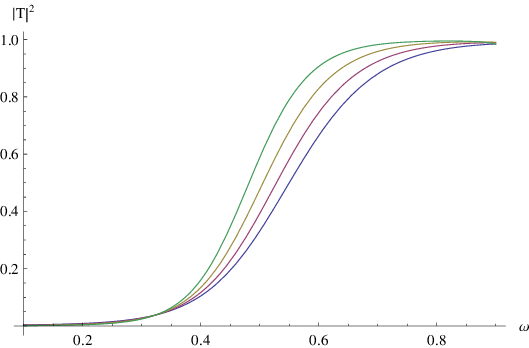}~~~\includegraphics{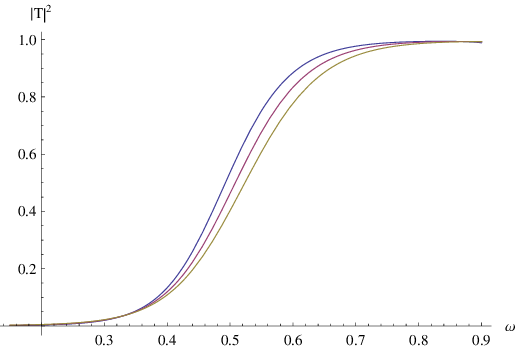}~~~\includegraphics{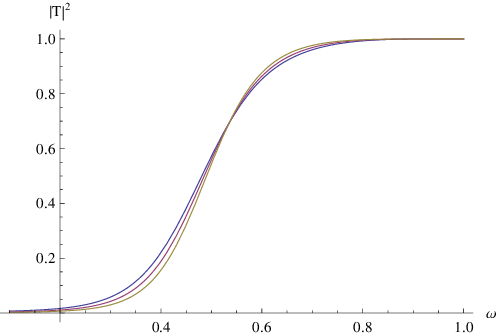}}
\caption{Transmission coefficients for $\ell=1$ electromagnetic perturbations: (left) $\epsilon=-0.2$, $-0.1,$ $0,$ $0.2$, $a_1 = 0$, $b_1 = 0$; (middle) $\epsilon=0$, $b_1 =0.1$, $a_1 =-0.2$, $0,$ $0.2$; (left) $\epsilon=0.1$, $a_1 =0.1$, $b_1 =-0.2$, $0,$ $0.2$.}\label{fig:L1s1par}
\end{figure*}
\begin{figure*}
\resizebox{\linewidth}{!}{\includegraphics{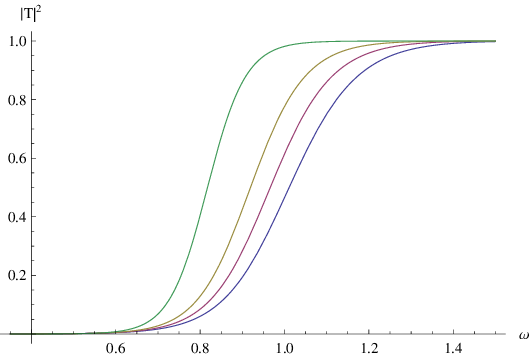}~~~\includegraphics{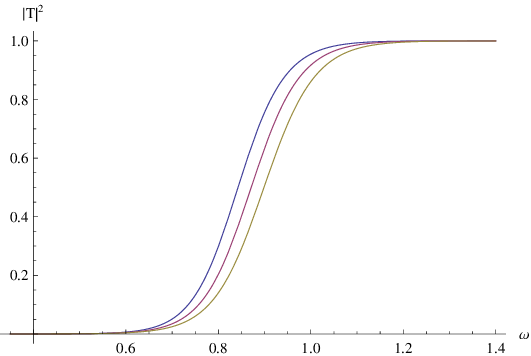}~~~\includegraphics{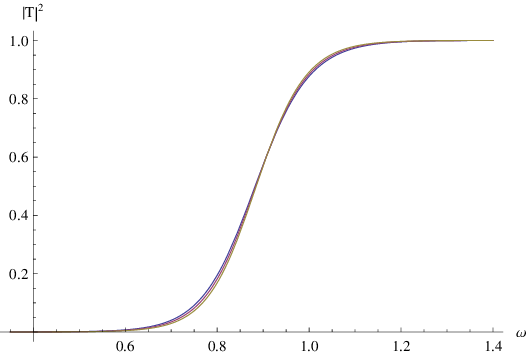}}
\caption{Transmission coefficients for $\ell=2$ electromagnetic perturbations: (left) $\epsilon=-0.2$, $-0.1,$ $0,$ $0.2$, $a_1 = 0$, $b_1 = 0$; (middle) $\epsilon=0.1$, $b_1 =0.1$, $a_1 =-0.2$, $0,$ $0.2$; (left) $\epsilon=0.1$, $a_1 =0.1$, $b_1 =-0.2$, $0,$ $0.2$.}\label{fig:L2s1par}
\end{figure*}

From Fig. \ref{fig:L2s0}, we observe that when \(\ell\) is greater than \(s\), the analytic formula provides results that are quite close to those obtained using the 6th-order WKB method. However, even for \(\ell = s\), the relative error can reach a few percent, as demonstrated in Fig. \ref{fig:L1s1} for \(s = \ell = 1\) perturbations. The situation is even worse for gravitational perturbations with \(\ell = 2\), where the relative error can reach up to 15\% even at relatively small values of the parametrization coefficients. However, extending the analytic formula to the next order in \(\kappa^{-1}\) reduces the relative error to just a few percent, as shown in fig. \ref{fig:L2s2}. We can also observe that the parameter \(\epsilon\) is the most dominant, with the grey-body factors being most sensitive to this parameter, while \(a_1\) and \(b_1\) are subdominant coefficients.

From Figs. \ref{fig:L1s1par} and \ref{fig:L2s1par}, we observe that the dominant parameter influencing the change in the grey-body factor is \(\epsilon\). As the deviation of the radius from its Schwarzschild limit increases from negative values, through zero, to positive values, the grey-body factors increase. This implies that the higher the compactness of the black hole, the higher the grey-body factors. In other words, the effective potentials around a more compact black hole (i.e., one that is heavier than the Schwarzschild black hole at the same fixed radius of the event horizon) reflect a smaller fraction of particles back to the event horizon. 

The coefficients \(a_1\) and \(b_2\), which determine the geometry in the near-horizon region, typically exhibit different behaviors at small and large frequencies and cause relatively small shifts in the grey-body factors. Generally, as \(a_1\) increases, the grey-body factors decrease unless the frequencies \(\omega\) are very low, where almost complete reflection occurs. The dependence on \(b_1\) is more complex and is more strongly influenced by the spin of the field under consideration.

We have observed that the higher-order coefficients of the parametrization, \(a_2\) and \(b_2\), result in only minor changes to the grey-body factors, even when their values are not particularly small and no expansion in terms of powers of \(a_i\) and \(b_i\) is employed for the calculations (see fig. \ref{fig:a2}). From this observation, an important conclusion can be drawn: the stability of the grey-body factors with respect to changes in the higher-order coefficients \(a_i\) and \(b_i\), which govern the near-horizon geometry, explains the robustness of the grey-body factors against near-horizon deformations of the geometry. These near-horizon deformations are the primary factor contributing to the high sensitivity of the overtones of quasinormal modes \cite{Konoplya:2022pbc}. The observed stability of the grey-body factors in general spherically symmetric black hole spacetimes is consistent with the stability observed in specific cases studied in \cite{Rosato:2024arw}.

\textbf{Conclusions.}
In this letter, we have analyzed the grey-body factors of general parametrized spherically symmetric black holes and demonstrated that the primary parameter influencing the grey-body factors is the deviation of the event horizon radius from its Schwarzschild limit. The coefficients of the parametrization that determine the near-horizon geometry are less significant, and their impact on the grey-body factors diminishes rapidly as the order increases. This indicates, in a general framework, that grey-body factors are much more stable characteristics than quasinormal frequencies, which aligns with recent observations \cite{Rosato:2024arw}. We have also derived an analytic formula for the grey-body factor of the parametrized black hole and shown that, for large and intermediate values of the multipole numbers, it is sufficiently accurate.

\textbf{Acknowledgments.}
Antonina F. Zinhailo acknowledges Silesian University grant SGS 2024 for support. Alexey Dubinsky acknowledges the Pablo de Olavide University for their support through the University-Refuge Action Plan. The authors are thankful to R. A. Konoplya for useful discussions.

\bibliography{Bibliography}

\begin{thebibliography}{76}%
\makeatletter
\providecommand \@ifxundefined [1]{%
 \@ifx{#1\undefined}
}%
\providecommand \@ifnum [1]{%
 \ifnum #1\expandafter \@firstoftwo
 \else \expandafter \@secondoftwo
 \fi
}%
\providecommand \@ifx [1]{%
 \ifx #1\expandafter \@firstoftwo
 \else \expandafter \@secondoftwo
 \fi
}%
\providecommand \natexlab [1]{#1}%
\providecommand \enquote  [1]{``#1''}%
\providecommand \bibnamefont  [1]{#1}%
\providecommand \bibfnamefont [1]{#1}%
\providecommand \citenamefont [1]{#1}%
\providecommand \href@noop [0]{\@secondoftwo}%
\providecommand \href [0]{\begingroup \@sanitize@url \@href}%
\providecommand \@href[1]{\@@startlink{#1}\@@href}%
\providecommand \@@href[1]{\endgroup#1\@@endlink}%
\providecommand \@sanitize@url [0]{\catcode `\\12\catcode `\$12\catcode `\&12\catcode `\#12\catcode `\^12\catcode `\_12\catcode `\%12\relax}%
\providecommand \@@startlink[1]{}%
\providecommand \@@endlink[0]{}%
\providecommand \url  [0]{\begingroup\@sanitize@url \@url }%
\providecommand \@url [1]{\endgroup\@href {#1}{\urlprefix }}%
\providecommand \urlprefix  [0]{URL }%
\providecommand \Eprint [0]{\href }%
\providecommand \doibase [0]{http://dx.doi.org/}%
\providecommand \selectlanguage [0]{\@gobble}%
\providecommand \bibinfo  [0]{\@secondoftwo}%
\providecommand \bibfield  [0]{\@secondoftwo}%
\providecommand \translation [1]{[#1]}%
\providecommand \BibitemOpen [0]{}%
\providecommand \bibitemStop [0]{}%
\providecommand \bibitemNoStop [0]{.\EOS\space}%
\providecommand \EOS [0]{\spacefactor3000\relax}%
\providecommand \BibitemShut  [1]{\csname bibitem#1\endcsname}%
\let\auto@bib@innerbib\@empty
\bibitem [{\citenamefont {Bekenstein}(1973)}]{Bekenstein:1973ur}%
  \BibitemOpen
  \bibfield  {author} {\bibinfo {author} {\bibfnamefont {J.~D.}\ \bibnamefont {Bekenstein}},\ }\href {\doibase 10.1103/PhysRevD.7.2333} {\bibfield  {journal} {\bibinfo  {journal} {Phys. Rev. D}\ }\textbf {\bibinfo {volume} {7}},\ \bibinfo {pages} {2333} (\bibinfo {year} {1973})}\BibitemShut {NoStop}%
\bibitem [{\citenamefont {Hawking}(1975)}]{Hawking:1975vcx}%
  \BibitemOpen
  \bibfield  {author} {\bibinfo {author} {\bibfnamefont {S.~W.}\ \bibnamefont {Hawking}},\ }\href {\doibase 10.1007/BF02345020} {\bibfield  {journal} {\bibinfo  {journal} {Commun. Math. Phys.}\ }\textbf {\bibinfo {volume} {43}},\ \bibinfo {pages} {199} (\bibinfo {year} {1975})},\ \bibinfo {note} {[Erratum: Commun.Math.Phys. 46, 206 (1976)]}\BibitemShut {NoStop}%
\bibitem [{\citenamefont {Oshita}(2024)}]{Oshita:2023cjz}%
  \BibitemOpen
  \bibfield  {author} {\bibinfo {author} {\bibfnamefont {N.}~\bibnamefont {Oshita}},\ }\href {\doibase 10.1103/PhysRevD.109.104028} {\bibfield  {journal} {\bibinfo  {journal} {Phys. Rev. D}\ }\textbf {\bibinfo {volume} {109}},\ \bibinfo {pages} {104028} (\bibinfo {year} {2024})},\ \Eprint {http://arxiv.org/abs/2309.05725} {arXiv:2309.05725 [gr-qc]} \BibitemShut {NoStop}%
\bibitem [{\citenamefont {Konoplya}\ and\ \citenamefont {Zhidenko}(2011)}]{Konoplya:2011qq}%
  \BibitemOpen
  \bibfield  {author} {\bibinfo {author} {\bibfnamefont {R.~A.}\ \bibnamefont {Konoplya}}\ and\ \bibinfo {author} {\bibfnamefont {A.}~\bibnamefont {Zhidenko}},\ }\href {\doibase 10.1103/RevModPhys.83.793} {\bibfield  {journal} {\bibinfo  {journal} {Rev. Mod. Phys.}\ }\textbf {\bibinfo {volume} {83}},\ \bibinfo {pages} {793} (\bibinfo {year} {2011})},\ \Eprint {http://arxiv.org/abs/1102.4014} {arXiv:1102.4014 [gr-qc]} \BibitemShut {NoStop}%
\bibitem [{\citenamefont {Kokkotas}\ and\ \citenamefont {Schmidt}(1999)}]{Kokkotas:1999bd}%
  \BibitemOpen
  \bibfield  {author} {\bibinfo {author} {\bibfnamefont {K.~D.}\ \bibnamefont {Kokkotas}}\ and\ \bibinfo {author} {\bibfnamefont {B.~G.}\ \bibnamefont {Schmidt}},\ }\href {\doibase 10.12942/lrr-1999-2} {\bibfield  {journal} {\bibinfo  {journal} {Living Rev. Rel.}\ }\textbf {\bibinfo {volume} {2}},\ \bibinfo {pages} {2} (\bibinfo {year} {1999})},\ \Eprint {http://arxiv.org/abs/gr-qc/9909058} {arXiv:gr-qc/9909058} \BibitemShut {NoStop}%
\bibitem [{\citenamefont {Konoplya}\ and\ \citenamefont {Zhidenko}(2022)}]{Konoplya:2022pbc}%
  \BibitemOpen
  \bibfield  {author} {\bibinfo {author} {\bibfnamefont {R.~A.}\ \bibnamefont {Konoplya}}\ and\ \bibinfo {author} {\bibfnamefont {A.}~\bibnamefont {Zhidenko}},\ }\href@noop {} {\bibfield  {journal} {\bibinfo  {journal} {arXiv: 2209.00679}\ } (\bibinfo {year} {2022})},\ \Eprint {http://arxiv.org/abs/2209.00679} {arXiv:2209.00679 [gr-qc]} \BibitemShut {NoStop}%
\bibitem [{\citenamefont {Rosato}\ \emph {et~al.}(2024)\citenamefont {Rosato}, \citenamefont {Destounis},\ and\ \citenamefont {Pani}}]{Rosato:2024arw}%
  \BibitemOpen
  \bibfield  {author} {\bibinfo {author} {\bibfnamefont {R.~F.}\ \bibnamefont {Rosato}}, \bibinfo {author} {\bibfnamefont {K.}~\bibnamefont {Destounis}}, \ and\ \bibinfo {author} {\bibfnamefont {P.}~\bibnamefont {Pani}},\ }\href@noop {} {\bibfield  {journal} {\bibinfo  {journal} {arXiv:2406.01692}\ } (\bibinfo {year} {2024})},\ \Eprint {http://arxiv.org/abs/2406.01692} {arXiv:2406.01692 [gr-qc]} \BibitemShut {NoStop}%
\bibitem [{\citenamefont {Oshita}\ \emph {et~al.}(2024)\citenamefont {Oshita}, \citenamefont {Takahashi},\ and\ \citenamefont {Mukohyama}}]{Oshita:2024fzf}%
  \BibitemOpen
  \bibfield  {author} {\bibinfo {author} {\bibfnamefont {N.}~\bibnamefont {Oshita}}, \bibinfo {author} {\bibfnamefont {K.}~\bibnamefont {Takahashi}}, \ and\ \bibinfo {author} {\bibfnamefont {S.}~\bibnamefont {Mukohyama}},\ }\href@noop {} {\bibfield  {journal} {\bibinfo  {journal} {arXiv:2406.04525}\ } (\bibinfo {year} {2024})},\ \Eprint {http://arxiv.org/abs/2406.04525} {arXiv:2406.04525 [gr-qc]} \BibitemShut {NoStop}%
\bibitem [{\citenamefont {Konoplya}\ and\ \citenamefont {Zhidenko}(2024{\natexlab{a}})}]{Konoplya:2024lir}%
  \BibitemOpen
  \bibfield  {author} {\bibinfo {author} {\bibfnamefont {R.~A.}\ \bibnamefont {Konoplya}}\ and\ \bibinfo {author} {\bibfnamefont {A.}~\bibnamefont {Zhidenko}},\ }\href {\doibase 10.1088/1475-7516/2024/09/068} {\bibfield  {journal} {\bibinfo  {journal} {JCAP}\ }\textbf {\bibinfo {volume} {09}},\ \bibinfo {pages} {068} (\bibinfo {year} {2024}{\natexlab{a}})},\ \Eprint {http://arxiv.org/abs/2406.11694} {arXiv:2406.11694 [gr-qc]} \BibitemShut {NoStop}%
\bibitem [{\citenamefont {Konoplya}\ and\ \citenamefont {Zhidenko}(2024{\natexlab{b}})}]{Konoplya:2024vuj}%
  \BibitemOpen
  \bibfield  {author} {\bibinfo {author} {\bibfnamefont {R.~A.}\ \bibnamefont {Konoplya}}\ and\ \bibinfo {author} {\bibfnamefont {A.}~\bibnamefont {Zhidenko}},\ }\href@noop {} {\  (\bibinfo {year} {2024}{\natexlab{b}})},\ \Eprint {http://arxiv.org/abs/2408.11162} {arXiv:2408.11162 [gr-qc]} \BibitemShut {NoStop}%
\bibitem [{\citenamefont {Oshita}\ and\ \citenamefont {Cardoso}(2024)}]{Oshita:2024wgt}%
  \BibitemOpen
  \bibfield  {author} {\bibinfo {author} {\bibfnamefont {N.}~\bibnamefont {Oshita}}\ and\ \bibinfo {author} {\bibfnamefont {V.}~\bibnamefont {Cardoso}},\ }\href@noop {} {\bibfield  {journal} {\bibinfo  {journal} {arXiv: 2407.02563}\ } (\bibinfo {year} {2024})},\ \Eprint {http://arxiv.org/abs/2407.02563} {arXiv:2407.02563 [gr-qc]} \BibitemShut {NoStop}%
\bibitem [{\citenamefont {Page}(1976{\natexlab{a}})}]{Page:1976df}%
  \BibitemOpen
  \bibfield  {author} {\bibinfo {author} {\bibfnamefont {D.~N.}\ \bibnamefont {Page}},\ }\href {\doibase 10.1103/PhysRevD.13.198} {\bibfield  {journal} {\bibinfo  {journal} {Phys. Rev. D}\ }\textbf {\bibinfo {volume} {13}},\ \bibinfo {pages} {198} (\bibinfo {year} {1976}{\natexlab{a}})}\BibitemShut {NoStop}%
\bibitem [{\citenamefont {Page}(1976{\natexlab{b}})}]{Page:1976ki}%
  \BibitemOpen
  \bibfield  {author} {\bibinfo {author} {\bibfnamefont {D.~N.}\ \bibnamefont {Page}},\ }\href {\doibase 10.1103/PhysRevD.14.3260} {\bibfield  {journal} {\bibinfo  {journal} {Phys. Rev. D}\ }\textbf {\bibinfo {volume} {14}},\ \bibinfo {pages} {3260} (\bibinfo {year} {1976}{\natexlab{b}})}\BibitemShut {NoStop}%
\bibitem [{\citenamefont {Page}(1977)}]{Page:1977um}%
  \BibitemOpen
  \bibfield  {author} {\bibinfo {author} {\bibfnamefont {D.~N.}\ \bibnamefont {Page}},\ }\href {\doibase 10.1103/PhysRevD.16.2402} {\bibfield  {journal} {\bibinfo  {journal} {Phys. Rev. D}\ }\textbf {\bibinfo {volume} {16}},\ \bibinfo {pages} {2402} (\bibinfo {year} {1977})}\BibitemShut {NoStop}%
\bibitem [{\citenamefont {Kanti}(2004)}]{Kanti:2004nr}%
  \BibitemOpen
  \bibfield  {author} {\bibinfo {author} {\bibfnamefont {P.}~\bibnamefont {Kanti}},\ }\href {\doibase 10.1142/S0217751X04018324} {\bibfield  {journal} {\bibinfo  {journal} {Int. J. Mod. Phys. A}\ }\textbf {\bibinfo {volume} {19}},\ \bibinfo {pages} {4899} (\bibinfo {year} {2004})},\ \Eprint {http://arxiv.org/abs/hep-ph/0402168} {arXiv:hep-ph/0402168} \BibitemShut {NoStop}%
\bibitem [{\citenamefont {Grain}\ \emph {et~al.}(2005)\citenamefont {Grain}, \citenamefont {Barrau},\ and\ \citenamefont {Kanti}}]{Grain:2005my}%
  \BibitemOpen
  \bibfield  {author} {\bibinfo {author} {\bibfnamefont {J.}~\bibnamefont {Grain}}, \bibinfo {author} {\bibfnamefont {A.}~\bibnamefont {Barrau}}, \ and\ \bibinfo {author} {\bibfnamefont {P.}~\bibnamefont {Kanti}},\ }\href {\doibase 10.1103/PhysRevD.72.104016} {\bibfield  {journal} {\bibinfo  {journal} {Phys. Rev. D}\ }\textbf {\bibinfo {volume} {72}},\ \bibinfo {pages} {104016} (\bibinfo {year} {2005})},\ \Eprint {http://arxiv.org/abs/hep-th/0509128} {arXiv:hep-th/0509128} \BibitemShut {NoStop}%
\bibitem [{\citenamefont {Kanti}\ \emph {et~al.}(2014)\citenamefont {Kanti}, \citenamefont {Pappas},\ and\ \citenamefont {Pappas}}]{Kanti:2014dxa}%
  \BibitemOpen
  \bibfield  {author} {\bibinfo {author} {\bibfnamefont {P.}~\bibnamefont {Kanti}}, \bibinfo {author} {\bibfnamefont {T.}~\bibnamefont {Pappas}}, \ and\ \bibinfo {author} {\bibfnamefont {N.}~\bibnamefont {Pappas}},\ }\href {\doibase 10.1103/PhysRevD.90.124077} {\bibfield  {journal} {\bibinfo  {journal} {Phys. Rev. D}\ }\textbf {\bibinfo {volume} {90}},\ \bibinfo {pages} {124077} (\bibinfo {year} {2014})},\ \Eprint {http://arxiv.org/abs/1409.8664} {arXiv:1409.8664 [hep-th]} \BibitemShut {NoStop}%
\bibitem [{\citenamefont {Pappas}\ \emph {et~al.}(2016)\citenamefont {Pappas}, \citenamefont {Kanti},\ and\ \citenamefont {Pappas}}]{Pappas:2016ovo}%
  \BibitemOpen
  \bibfield  {author} {\bibinfo {author} {\bibfnamefont {T.}~\bibnamefont {Pappas}}, \bibinfo {author} {\bibfnamefont {P.}~\bibnamefont {Kanti}}, \ and\ \bibinfo {author} {\bibfnamefont {N.}~\bibnamefont {Pappas}},\ }\href {\doibase 10.1103/PhysRevD.94.024035} {\bibfield  {journal} {\bibinfo  {journal} {Phys. Rev. D}\ }\textbf {\bibinfo {volume} {94}},\ \bibinfo {pages} {024035} (\bibinfo {year} {2016})},\ \Eprint {http://arxiv.org/abs/1604.08617} {arXiv:1604.08617 [hep-th]} \BibitemShut {NoStop}%
\bibitem [{\citenamefont {Rezzolla}\ and\ \citenamefont {Zhidenko}(2014)}]{Rezzolla:2014mua}%
  \BibitemOpen
  \bibfield  {author} {\bibinfo {author} {\bibfnamefont {L.}~\bibnamefont {Rezzolla}}\ and\ \bibinfo {author} {\bibfnamefont {A.}~\bibnamefont {Zhidenko}},\ }\href {\doibase 10.1103/PhysRevD.90.084009} {\bibfield  {journal} {\bibinfo  {journal} {Phys. Rev. D}\ }\textbf {\bibinfo {volume} {90}},\ \bibinfo {pages} {084009} (\bibinfo {year} {2014})},\ \Eprint {http://arxiv.org/abs/1407.3086} {arXiv:1407.3086 [gr-qc]} \BibitemShut {NoStop}%
\bibitem [{\citenamefont {Konoplya}\ and\ \citenamefont {Zhidenko}(2020)}]{Konoplya:2020hyk}%
  \BibitemOpen
  \bibfield  {author} {\bibinfo {author} {\bibfnamefont {R.~A.}\ \bibnamefont {Konoplya}}\ and\ \bibinfo {author} {\bibfnamefont {A.}~\bibnamefont {Zhidenko}},\ }\href {\doibase 10.1103/PhysRevD.101.124004} {\bibfield  {journal} {\bibinfo  {journal} {Phys. Rev. D}\ }\textbf {\bibinfo {volume} {101}},\ \bibinfo {pages} {124004} (\bibinfo {year} {2020})},\ \Eprint {http://arxiv.org/abs/2001.06100} {arXiv:2001.06100 [gr-qc]} \BibitemShut {NoStop}%
\bibitem [{\citenamefont {Konoplya}\ and\ \citenamefont {Zhidenko}(2021)}]{Konoplya:2021slg}%
  \BibitemOpen
  \bibfield  {author} {\bibinfo {author} {\bibfnamefont {R.~A.}\ \bibnamefont {Konoplya}}\ and\ \bibinfo {author} {\bibfnamefont {A.}~\bibnamefont {Zhidenko}},\ }\href {\doibase 10.1103/PhysRevD.103.104033} {\bibfield  {journal} {\bibinfo  {journal} {Phys. Rev. D}\ }\textbf {\bibinfo {volume} {103}},\ \bibinfo {pages} {104033} (\bibinfo {year} {2021})},\ \Eprint {http://arxiv.org/abs/2103.03855} {arXiv:2103.03855 [gr-qc]} \BibitemShut {NoStop}%
\bibitem [{\citenamefont {Konoplya}\ and\ \citenamefont {Zhidenko}(2010)}]{Konoplya:2010kv}%
  \BibitemOpen
  \bibfield  {author} {\bibinfo {author} {\bibfnamefont {R.~A.}\ \bibnamefont {Konoplya}}\ and\ \bibinfo {author} {\bibfnamefont {A.}~\bibnamefont {Zhidenko}},\ }\href {\doibase 10.1103/PhysRevD.81.124036} {\bibfield  {journal} {\bibinfo  {journal} {Phys. Rev. D}\ }\textbf {\bibinfo {volume} {81}},\ \bibinfo {pages} {124036} (\bibinfo {year} {2010})},\ \Eprint {http://arxiv.org/abs/1004.1284} {arXiv:1004.1284 [hep-th]} \BibitemShut {NoStop}%
\bibitem [{\citenamefont {Kokkotas}\ \emph {et~al.}(2011)\citenamefont {Kokkotas}, \citenamefont {Konoplya},\ and\ \citenamefont {Zhidenko}}]{Kokkotas:2010zd}%
  \BibitemOpen
  \bibfield  {author} {\bibinfo {author} {\bibfnamefont {K.~D.}\ \bibnamefont {Kokkotas}}, \bibinfo {author} {\bibfnamefont {R.~A.}\ \bibnamefont {Konoplya}}, \ and\ \bibinfo {author} {\bibfnamefont {A.}~\bibnamefont {Zhidenko}},\ }\href {\doibase 10.1103/PhysRevD.83.024031} {\bibfield  {journal} {\bibinfo  {journal} {Phys. Rev. D}\ }\textbf {\bibinfo {volume} {83}},\ \bibinfo {pages} {024031} (\bibinfo {year} {2011})},\ \Eprint {http://arxiv.org/abs/1011.1843} {arXiv:1011.1843 [gr-qc]} \BibitemShut {NoStop}%
\bibitem [{\citenamefont {Konoplya}\ and\ \citenamefont {Zhidenko}(2023)}]{Konoplya:2023moy}%
  \BibitemOpen
  \bibfield  {author} {\bibinfo {author} {\bibfnamefont {R.~A.}\ \bibnamefont {Konoplya}}\ and\ \bibinfo {author} {\bibfnamefont {A.}~\bibnamefont {Zhidenko}},\ }\href {\doibase 10.1088/1361-6382/ad0a52} {\bibfield  {journal} {\bibinfo  {journal} {Class. Quant. Grav.}\ }\textbf {\bibinfo {volume} {40}},\ \bibinfo {pages} {245005} (\bibinfo {year} {2023})},\ \Eprint {http://arxiv.org/abs/2309.02560} {arXiv:2309.02560 [gr-qc]} \BibitemShut {NoStop}%
\bibitem [{\citenamefont {Malik}(2024{\natexlab{a}})}]{Malik:2024sxv}%
  \BibitemOpen
  \bibfield  {author} {\bibinfo {author} {\bibfnamefont {Z.}~\bibnamefont {Malik}},\ }\href {\doibase 10.1007/s10773-024-05660-5} {\bibfield  {journal} {\bibinfo  {journal} {Int. J. Theor. Phys.}\ }\textbf {\bibinfo {volume} {63}},\ \bibinfo {pages} {128} (\bibinfo {year} {2024}{\natexlab{a}})}\BibitemShut {NoStop}%
\bibitem [{\citenamefont {Malik}(2024{\natexlab{b}})}]{Malik:2024voy}%
  \BibitemOpen
  \bibfield  {author} {\bibinfo {author} {\bibfnamefont {Z.}~\bibnamefont {Malik}},\ }\href {\doibase 10.1142/S0217751X24500246} {\bibfield  {journal} {\bibinfo  {journal} {Int. J. Mod. Phys. A}\ }\textbf {\bibinfo {volume} {39}},\ \bibinfo {pages} {2450024} (\bibinfo {year} {2024}{\natexlab{b}})}\BibitemShut {NoStop}%
\bibitem [{\citenamefont {Bolokhov}(2024{\natexlab{a}})}]{Bolokhov:2024ixe}%
  \BibitemOpen
  \bibfield  {author} {\bibinfo {author} {\bibfnamefont {S.~V.}\ \bibnamefont {Bolokhov}},\ }\href {\doibase 10.1140/epjc/s10052-024-12990-5} {\bibfield  {journal} {\bibinfo  {journal} {Eur. Phys. J. C}\ }\textbf {\bibinfo {volume} {84}},\ \bibinfo {pages} {634} (\bibinfo {year} {2024}{\natexlab{a}})},\ \Eprint {http://arxiv.org/abs/2404.09364} {arXiv:2404.09364 [gr-qc]} \BibitemShut {NoStop}%
\bibitem [{\citenamefont {Malik}(2024{\natexlab{c}})}]{Malik:2024tuf}%
  \BibitemOpen
  \bibfield  {author} {\bibinfo {author} {\bibfnamefont {Z.}~\bibnamefont {Malik}},\ }\href {\doibase 10.1209/0295-5075/ad7885} {\bibfield  {journal} {\bibinfo  {journal} {EPL}\ }\textbf {\bibinfo {volume} {147}},\ \bibinfo {pages} {69001} (\bibinfo {year} {2024}{\natexlab{c}})},\ \Eprint {http://arxiv.org/abs/2410.04306} {arXiv:2410.04306 [gr-qc]} \BibitemShut {NoStop}%
\bibitem [{\citenamefont {Malik}(2024{\natexlab{d}})}]{2753764}%
  \BibitemOpen
  \bibfield  {author} {\bibinfo {author} {\bibfnamefont {Z.}~\bibnamefont {Malik}},\ }\href {\doibase 10.13140/RG.2.2.27879.83363} {\  (\bibinfo {year} {2024}{\natexlab{d}}),\ 10.13140/RG.2.2.27879.83363}\BibitemShut {NoStop}%
\bibitem [{\citenamefont {Malik}(2024{\natexlab{e}})}]{Malik:2024nhy}%
  \BibitemOpen
  \bibfield  {author} {\bibinfo {author} {\bibfnamefont {Z.}~\bibnamefont {Malik}},\ }\href@noop {} {\  (\bibinfo {year} {2024}{\natexlab{e}})},\ \Eprint {http://arxiv.org/abs/2409.01561} {arXiv:2409.01561 [gr-qc]} \BibitemShut {NoStop}%
\bibitem [{\citenamefont {Konoplya}\ \emph {et~al.}(2016)\citenamefont {Konoplya}, \citenamefont {Rezzolla},\ and\ \citenamefont {Zhidenko}}]{Konoplya:2016jvv}%
  \BibitemOpen
  \bibfield  {author} {\bibinfo {author} {\bibfnamefont {R.}~\bibnamefont {Konoplya}}, \bibinfo {author} {\bibfnamefont {L.}~\bibnamefont {Rezzolla}}, \ and\ \bibinfo {author} {\bibfnamefont {A.}~\bibnamefont {Zhidenko}},\ }\href {\doibase 10.1103/PhysRevD.93.064015} {\bibfield  {journal} {\bibinfo  {journal} {Phys. Rev. D}\ }\textbf {\bibinfo {volume} {93}},\ \bibinfo {pages} {064015} (\bibinfo {year} {2016})},\ \Eprint {http://arxiv.org/abs/1602.02378} {arXiv:1602.02378 [gr-qc]} \BibitemShut {NoStop}%
\bibitem [{\citenamefont {Kocherlakota}\ and\ \citenamefont {Rezzolla}(2020)}]{Kocherlakota:2020kyu}%
  \BibitemOpen
  \bibfield  {author} {\bibinfo {author} {\bibfnamefont {P.}~\bibnamefont {Kocherlakota}}\ and\ \bibinfo {author} {\bibfnamefont {L.}~\bibnamefont {Rezzolla}},\ }\href {\doibase 10.1103/PhysRevD.102.064058} {\bibfield  {journal} {\bibinfo  {journal} {Phys. Rev. D}\ }\textbf {\bibinfo {volume} {102}},\ \bibinfo {pages} {064058} (\bibinfo {year} {2020})},\ \Eprint {http://arxiv.org/abs/2007.15593} {arXiv:2007.15593 [gr-qc]} \BibitemShut {NoStop}%
\bibitem [{\citenamefont {Zhang}(2024)}]{Zhang:2024rvk}%
  \BibitemOpen
  \bibfield  {author} {\bibinfo {author} {\bibfnamefont {S.-J.}\ \bibnamefont {Zhang}},\ }\href {\doibase 10.1103/PhysRevD.109.084066} {\bibfield  {journal} {\bibinfo  {journal} {Phys. Rev. D}\ }\textbf {\bibinfo {volume} {109}},\ \bibinfo {pages} {084066} (\bibinfo {year} {2024})},\ \Eprint {http://arxiv.org/abs/2402.15050} {arXiv:2402.15050 [gr-qc]} \BibitemShut {NoStop}%
\bibitem [{\citenamefont {Cassing}\ and\ \citenamefont {Rezzolla}(2023)}]{Cassing:2023bpt}%
  \BibitemOpen
  \bibfield  {author} {\bibinfo {author} {\bibfnamefont {M.}~\bibnamefont {Cassing}}\ and\ \bibinfo {author} {\bibfnamefont {L.}~\bibnamefont {Rezzolla}},\ }\href {\doibase 10.1093/mnras/stad1039} {\bibfield  {journal} {\bibinfo  {journal} {Mon. Not. Roy. Astron. Soc.}\ }\textbf {\bibinfo {volume} {522}},\ \bibinfo {pages} {2415} (\bibinfo {year} {2023})},\ \Eprint {http://arxiv.org/abs/2302.09135} {arXiv:2302.09135 [gr-qc]} \BibitemShut {NoStop}%
\bibitem [{\citenamefont {Li}\ \emph {et~al.}(2021)\citenamefont {Li}, \citenamefont {Abdujabbarov},\ and\ \citenamefont {Han}}]{Li:2021mnx}%
  \BibitemOpen
  \bibfield  {author} {\bibinfo {author} {\bibfnamefont {S.}~\bibnamefont {Li}}, \bibinfo {author} {\bibfnamefont {A.~A.}\ \bibnamefont {Abdujabbarov}}, \ and\ \bibinfo {author} {\bibfnamefont {W.-B.}\ \bibnamefont {Han}},\ }\href {\doibase 10.1140/epjc/s10052-021-09445-6} {\bibfield  {journal} {\bibinfo  {journal} {Eur. Phys. J. C}\ }\textbf {\bibinfo {volume} {81}},\ \bibinfo {pages} {649} (\bibinfo {year} {2021})},\ \Eprint {http://arxiv.org/abs/2103.08104} {arXiv:2103.08104 [gr-qc]} \BibitemShut {NoStop}%
\bibitem [{\citenamefont {Ma}\ and\ \citenamefont {Rezzolla}(2024)}]{Ma:2024kbu}%
  \BibitemOpen
  \bibfield  {author} {\bibinfo {author} {\bibfnamefont {Y.}~\bibnamefont {Ma}}\ and\ \bibinfo {author} {\bibfnamefont {L.}~\bibnamefont {Rezzolla}},\ }\href {\doibase 10.1103/PhysRevD.110.024032} {\bibfield  {journal} {\bibinfo  {journal} {Phys. Rev. D}\ }\textbf {\bibinfo {volume} {110}},\ \bibinfo {pages} {024032} (\bibinfo {year} {2024})},\ \Eprint {http://arxiv.org/abs/2404.06509} {arXiv:2404.06509 [gr-qc]} \BibitemShut {NoStop}%
\bibitem [{\citenamefont {Bronnikov}\ \emph {et~al.}(2021)\citenamefont {Bronnikov}, \citenamefont {Konoplya},\ and\ \citenamefont {Pappas}}]{Bronnikov:2021liv}%
  \BibitemOpen
  \bibfield  {author} {\bibinfo {author} {\bibfnamefont {K.~A.}\ \bibnamefont {Bronnikov}}, \bibinfo {author} {\bibfnamefont {R.~A.}\ \bibnamefont {Konoplya}}, \ and\ \bibinfo {author} {\bibfnamefont {T.~D.}\ \bibnamefont {Pappas}},\ }\href {\doibase 10.1103/PhysRevD.103.124062} {\bibfield  {journal} {\bibinfo  {journal} {Phys. Rev. D}\ }\textbf {\bibinfo {volume} {103}},\ \bibinfo {pages} {124062} (\bibinfo {year} {2021})},\ \Eprint {http://arxiv.org/abs/2102.10679} {arXiv:2102.10679 [gr-qc]} \BibitemShut {NoStop}%
\bibitem [{\citenamefont {Shashank}\ and\ \citenamefont {Bambi}(2022)}]{Shashank:2021giy}%
  \BibitemOpen
  \bibfield  {author} {\bibinfo {author} {\bibfnamefont {S.}~\bibnamefont {Shashank}}\ and\ \bibinfo {author} {\bibfnamefont {C.}~\bibnamefont {Bambi}},\ }\href {\doibase 10.1103/PhysRevD.105.104004} {\bibfield  {journal} {\bibinfo  {journal} {Phys. Rev. D}\ }\textbf {\bibinfo {volume} {105}},\ \bibinfo {pages} {104004} (\bibinfo {year} {2022})},\ \Eprint {http://arxiv.org/abs/2112.05388} {arXiv:2112.05388 [gr-qc]} \BibitemShut {NoStop}%
\bibitem [{\citenamefont {Kokkotas}\ \emph {et~al.}(2017)\citenamefont {Kokkotas}, \citenamefont {Konoplya},\ and\ \citenamefont {Zhidenko}}]{Kokkotas:2017zwt}%
  \BibitemOpen
  \bibfield  {author} {\bibinfo {author} {\bibfnamefont {K.}~\bibnamefont {Kokkotas}}, \bibinfo {author} {\bibfnamefont {R.~A.}\ \bibnamefont {Konoplya}}, \ and\ \bibinfo {author} {\bibfnamefont {A.}~\bibnamefont {Zhidenko}},\ }\href {\doibase 10.1103/PhysRevD.96.064007} {\bibfield  {journal} {\bibinfo  {journal} {Phys. Rev. D}\ }\textbf {\bibinfo {volume} {96}},\ \bibinfo {pages} {064007} (\bibinfo {year} {2017})},\ \Eprint {http://arxiv.org/abs/1705.09875} {arXiv:1705.09875 [gr-qc]} \BibitemShut {NoStop}%
\bibitem [{\citenamefont {Yu}\ \emph {et~al.}(2021)\citenamefont {Yu}, \citenamefont {Jiang}, \citenamefont {Abdikamalov}, \citenamefont {Ayzenberg}, \citenamefont {Bambi}, \citenamefont {Liu}, \citenamefont {Nampalliwar},\ and\ \citenamefont {Tripathi}}]{Yu:2021xen}%
  \BibitemOpen
  \bibfield  {author} {\bibinfo {author} {\bibfnamefont {Z.}~\bibnamefont {Yu}}, \bibinfo {author} {\bibfnamefont {Q.}~\bibnamefont {Jiang}}, \bibinfo {author} {\bibfnamefont {A.~B.}\ \bibnamefont {Abdikamalov}}, \bibinfo {author} {\bibfnamefont {D.}~\bibnamefont {Ayzenberg}}, \bibinfo {author} {\bibfnamefont {C.}~\bibnamefont {Bambi}}, \bibinfo {author} {\bibfnamefont {H.}~\bibnamefont {Liu}}, \bibinfo {author} {\bibfnamefont {S.}~\bibnamefont {Nampalliwar}}, \ and\ \bibinfo {author} {\bibfnamefont {A.}~\bibnamefont {Tripathi}},\ }\href {\doibase 10.1103/PhysRevD.104.084035} {\bibfield  {journal} {\bibinfo  {journal} {Phys. Rev. D}\ }\textbf {\bibinfo {volume} {104}},\ \bibinfo {pages} {084035} (\bibinfo {year} {2021})},\ \Eprint {http://arxiv.org/abs/2106.11658} {arXiv:2106.11658 [astro-ph.HE]} \BibitemShut {NoStop}%
\bibitem [{\citenamefont {Toshmatov}\ and\ \citenamefont {Ahmedov}(2023)}]{Toshmatov:2023anz}%
  \BibitemOpen
  \bibfield  {author} {\bibinfo {author} {\bibfnamefont {B.}~\bibnamefont {Toshmatov}}\ and\ \bibinfo {author} {\bibfnamefont {B.}~\bibnamefont {Ahmedov}},\ }\href {\doibase 10.1103/PhysRevD.108.084035} {\bibfield  {journal} {\bibinfo  {journal} {Phys. Rev. D}\ }\textbf {\bibinfo {volume} {108}},\ \bibinfo {pages} {084035} (\bibinfo {year} {2023})},\ \Eprint {http://arxiv.org/abs/2311.04602} {arXiv:2311.04602 [gr-qc]} \BibitemShut {NoStop}%
\bibitem [{\citenamefont {Nampalliwar}\ \emph {et~al.}(2020)\citenamefont {Nampalliwar}, \citenamefont {Xin}, \citenamefont {Srivastava}, \citenamefont {Abdikamalov}, \citenamefont {Ayzenberg}, \citenamefont {Bambi}, \citenamefont {Dauser}, \citenamefont {Garcia},\ and\ \citenamefont {Tripathi}}]{Nampalliwar:2019iti}%
  \BibitemOpen
  \bibfield  {author} {\bibinfo {author} {\bibfnamefont {S.}~\bibnamefont {Nampalliwar}}, \bibinfo {author} {\bibfnamefont {S.}~\bibnamefont {Xin}}, \bibinfo {author} {\bibfnamefont {S.}~\bibnamefont {Srivastava}}, \bibinfo {author} {\bibfnamefont {A.~B.}\ \bibnamefont {Abdikamalov}}, \bibinfo {author} {\bibfnamefont {D.}~\bibnamefont {Ayzenberg}}, \bibinfo {author} {\bibfnamefont {C.}~\bibnamefont {Bambi}}, \bibinfo {author} {\bibfnamefont {T.}~\bibnamefont {Dauser}}, \bibinfo {author} {\bibfnamefont {J.~A.}\ \bibnamefont {Garcia}}, \ and\ \bibinfo {author} {\bibfnamefont {A.}~\bibnamefont {Tripathi}},\ }\href {\doibase 10.1103/PhysRevD.102.124071} {\bibfield  {journal} {\bibinfo  {journal} {Phys. Rev. D}\ }\textbf {\bibinfo {volume} {102}},\ \bibinfo {pages} {124071} (\bibinfo {year} {2020})},\ \Eprint {http://arxiv.org/abs/1903.12119} {arXiv:1903.12119 [gr-qc]} \BibitemShut {NoStop}%
\bibitem [{\citenamefont {Ni}\ \emph {et~al.}(2016)\citenamefont {Ni}, \citenamefont {Jiang},\ and\ \citenamefont {Bambi}}]{Ni:2016uik}%
  \BibitemOpen
  \bibfield  {author} {\bibinfo {author} {\bibfnamefont {Y.}~\bibnamefont {Ni}}, \bibinfo {author} {\bibfnamefont {J.}~\bibnamefont {Jiang}}, \ and\ \bibinfo {author} {\bibfnamefont {C.}~\bibnamefont {Bambi}},\ }\href {\doibase 10.1088/1475-7516/2016/09/014} {\bibfield  {journal} {\bibinfo  {journal} {JCAP}\ }\textbf {\bibinfo {volume} {09}},\ \bibinfo {pages} {014} (\bibinfo {year} {2016})},\ \Eprint {http://arxiv.org/abs/1607.04893} {arXiv:1607.04893 [gr-qc]} \BibitemShut {NoStop}%
\bibitem [{\citenamefont {Konoplya}\ \emph {et~al.}(2018)\citenamefont {Konoplya}, \citenamefont {Stuchl\'\i{}k},\ and\ \citenamefont {Zhidenko}}]{Konoplya:2018arm}%
  \BibitemOpen
  \bibfield  {author} {\bibinfo {author} {\bibfnamefont {R.~A.}\ \bibnamefont {Konoplya}}, \bibinfo {author} {\bibfnamefont {Z.}~\bibnamefont {Stuchl\'\i{}k}}, \ and\ \bibinfo {author} {\bibfnamefont {A.}~\bibnamefont {Zhidenko}},\ }\href {\doibase 10.1103/PhysRevD.97.084044} {\bibfield  {journal} {\bibinfo  {journal} {Phys. Rev. D}\ }\textbf {\bibinfo {volume} {97}},\ \bibinfo {pages} {084044} (\bibinfo {year} {2018})},\ \Eprint {http://arxiv.org/abs/1801.07195} {arXiv:1801.07195 [gr-qc]} \BibitemShut {NoStop}%
\bibitem [{\citenamefont {Zinhailo}(2018)}]{Zinhailo:2018ska}%
  \BibitemOpen
  \bibfield  {author} {\bibinfo {author} {\bibfnamefont {A.~F.}\ \bibnamefont {Zinhailo}},\ }\href {\doibase 10.1140/epjc/s10052-018-6467-8} {\bibfield  {journal} {\bibinfo  {journal} {Eur. Phys. J. C}\ }\textbf {\bibinfo {volume} {78}},\ \bibinfo {pages} {992} (\bibinfo {year} {2018})},\ \Eprint {http://arxiv.org/abs/1809.03913} {arXiv:1809.03913 [gr-qc]} \BibitemShut {NoStop}%
\bibitem [{\citenamefont {Paul}(2024)}]{Paul:2023eep}%
  \BibitemOpen
  \bibfield  {author} {\bibinfo {author} {\bibfnamefont {P.}~\bibnamefont {Paul}},\ }\href {\doibase 10.1140/epjc/s10052-024-12563-6} {\bibfield  {journal} {\bibinfo  {journal} {Eur. Phys. J. C}\ }\textbf {\bibinfo {volume} {84}},\ \bibinfo {pages} {218} (\bibinfo {year} {2024})},\ \Eprint {http://arxiv.org/abs/2312.16479} {arXiv:2312.16479 [gr-qc]} \BibitemShut {NoStop}%
\bibitem [{\citenamefont {Abdalla}\ \emph {et~al.}(2006)\citenamefont {Abdalla}, \citenamefont {Cuadros-Melgar}, \citenamefont {Pavan},\ and\ \citenamefont {Molina}}]{Abdalla:2006qj}%
  \BibitemOpen
  \bibfield  {author} {\bibinfo {author} {\bibfnamefont {E.}~\bibnamefont {Abdalla}}, \bibinfo {author} {\bibfnamefont {B.}~\bibnamefont {Cuadros-Melgar}}, \bibinfo {author} {\bibfnamefont {A.~B.}\ \bibnamefont {Pavan}}, \ and\ \bibinfo {author} {\bibfnamefont {C.}~\bibnamefont {Molina}},\ }\href {\doibase 10.1016/j.nuclphysb.2006.06.017} {\bibfield  {journal} {\bibinfo  {journal} {Nucl. Phys. B}\ }\textbf {\bibinfo {volume} {752}},\ \bibinfo {pages} {40} (\bibinfo {year} {2006})},\ \Eprint {http://arxiv.org/abs/gr-qc/0604033} {arXiv:gr-qc/0604033} \BibitemShut {NoStop}%
\bibitem [{\citenamefont {Ashtekar}\ \emph {et~al.}(2018{\natexlab{a}})\citenamefont {Ashtekar}, \citenamefont {Olmedo},\ and\ \citenamefont {Singh}}]{Ashtekar:2018lag}%
  \BibitemOpen
  \bibfield  {author} {\bibinfo {author} {\bibfnamefont {A.}~\bibnamefont {Ashtekar}}, \bibinfo {author} {\bibfnamefont {J.}~\bibnamefont {Olmedo}}, \ and\ \bibinfo {author} {\bibfnamefont {P.}~\bibnamefont {Singh}},\ }\href {\doibase 10.1103/PhysRevLett.121.241301} {\bibfield  {journal} {\bibinfo  {journal} {Phys. Rev. Lett.}\ }\textbf {\bibinfo {volume} {121}},\ \bibinfo {pages} {241301} (\bibinfo {year} {2018}{\natexlab{a}})},\ \Eprint {http://arxiv.org/abs/1806.00648} {arXiv:1806.00648 [gr-qc]} \BibitemShut {NoStop}%
\bibitem [{\citenamefont {Ashtekar}\ \emph {et~al.}(2018{\natexlab{b}})\citenamefont {Ashtekar}, \citenamefont {Olmedo},\ and\ \citenamefont {Singh}}]{Ashtekar:2018cay}%
  \BibitemOpen
  \bibfield  {author} {\bibinfo {author} {\bibfnamefont {A.}~\bibnamefont {Ashtekar}}, \bibinfo {author} {\bibfnamefont {J.}~\bibnamefont {Olmedo}}, \ and\ \bibinfo {author} {\bibfnamefont {P.}~\bibnamefont {Singh}},\ }\href {\doibase 10.1103/PhysRevD.98.126003} {\bibfield  {journal} {\bibinfo  {journal} {Phys. Rev. D}\ }\textbf {\bibinfo {volume} {98}},\ \bibinfo {pages} {126003} (\bibinfo {year} {2018}{\natexlab{b}})},\ \Eprint {http://arxiv.org/abs/1806.02406} {arXiv:1806.02406 [gr-qc]} \BibitemShut {NoStop}%
\bibitem [{\citenamefont {Dehnen}\ and\ \citenamefont {Ivashchuk}(2004)}]{Dehnen:2003rc}%
  \BibitemOpen
  \bibfield  {author} {\bibinfo {author} {\bibfnamefont {H.}~\bibnamefont {Dehnen}}\ and\ \bibinfo {author} {\bibfnamefont {V.~D.}\ \bibnamefont {Ivashchuk}},\ }\href {\doibase 10.1063/1.1812357} {\bibfield  {journal} {\bibinfo  {journal} {J. Math. Phys.}\ }\textbf {\bibinfo {volume} {45}},\ \bibinfo {pages} {4726} (\bibinfo {year} {2004})},\ \Eprint {http://arxiv.org/abs/gr-qc/0310043} {arXiv:gr-qc/0310043} \BibitemShut {NoStop}%
\bibitem [{\citenamefont {Ivashchuk}(2010)}]{Ivashchuk:2010sn}%
  \BibitemOpen
  \bibfield  {author} {\bibinfo {author} {\bibfnamefont {V.~D.}\ \bibnamefont {Ivashchuk}},\ }\href {\doibase 10.1016/j.physletb.2010.08.060} {\bibfield  {journal} {\bibinfo  {journal} {Phys. Lett. B}\ }\textbf {\bibinfo {volume} {693}},\ \bibinfo {pages} {399} (\bibinfo {year} {2010})},\ \Eprint {http://arxiv.org/abs/1001.4053} {arXiv:1001.4053 [gr-qc]} \BibitemShut {NoStop}%
\bibitem [{\citenamefont {Bouhmadi-L\'opez}\ \emph {et~al.}(2020)\citenamefont {Bouhmadi-L\'opez}, \citenamefont {Brahma}, \citenamefont {Chen}, \citenamefont {Chen},\ and\ \citenamefont {Yeom}}]{Bouhmadi-Lopez:2020oia}%
  \BibitemOpen
  \bibfield  {author} {\bibinfo {author} {\bibfnamefont {M.}~\bibnamefont {Bouhmadi-L\'opez}}, \bibinfo {author} {\bibfnamefont {S.}~\bibnamefont {Brahma}}, \bibinfo {author} {\bibfnamefont {C.-Y.}\ \bibnamefont {Chen}}, \bibinfo {author} {\bibfnamefont {P.}~\bibnamefont {Chen}}, \ and\ \bibinfo {author} {\bibfnamefont {D.-h.}\ \bibnamefont {Yeom}},\ }\href {\doibase 10.1088/1475-7516/2020/07/066} {\bibfield  {journal} {\bibinfo  {journal} {JCAP}\ }\textbf {\bibinfo {volume} {07}},\ \bibinfo {pages} {066} (\bibinfo {year} {2020})},\ \Eprint {http://arxiv.org/abs/2004.13061} {arXiv:2004.13061 [gr-qc]} \BibitemShut {NoStop}%
\bibitem [{\citenamefont {Konoplya}\ and\ \citenamefont {Stashko}(2024)}]{Konoplya:2024lch}%
  \BibitemOpen
  \bibfield  {author} {\bibinfo {author} {\bibfnamefont {R.~A.}\ \bibnamefont {Konoplya}}\ and\ \bibinfo {author} {\bibfnamefont {O.~S.}\ \bibnamefont {Stashko}},\ }\href@noop {} {\  (\bibinfo {year} {2024})},\ \Eprint {http://arxiv.org/abs/2408.02578} {arXiv:2408.02578 [gr-qc]} \BibitemShut {NoStop}%
\bibitem [{\citenamefont {Dubinsky}(2024{\natexlab{a}})}]{Dubinsky:2024rvf}%
  \BibitemOpen
  \bibfield  {author} {\bibinfo {author} {\bibfnamefont {A.}~\bibnamefont {Dubinsky}},\ }\href@noop {} {\  (\bibinfo {year} {2024}{\natexlab{a}})},\ \Eprint {http://arxiv.org/abs/2409.16569} {arXiv:2409.16569 [gr-qc]} \BibitemShut {NoStop}%
\bibitem [{\citenamefont {Matyjasek}\ \emph {et~al.}(2024)\citenamefont {Matyjasek}, \citenamefont {Benda},\ and\ \citenamefont {Stafi\'nska}}]{Matyjasek:2024uwo}%
  \BibitemOpen
  \bibfield  {author} {\bibinfo {author} {\bibfnamefont {J.}~\bibnamefont {Matyjasek}}, \bibinfo {author} {\bibfnamefont {K.}~\bibnamefont {Benda}}, \ and\ \bibinfo {author} {\bibfnamefont {M.}~\bibnamefont {Stafi\'nska}},\ }\href {\doibase 10.1103/PhysRevD.110.064083} {\bibfield  {journal} {\bibinfo  {journal} {Phys. Rev. D}\ }\textbf {\bibinfo {volume} {110}},\ \bibinfo {pages} {064083} (\bibinfo {year} {2024})},\ \Eprint {http://arxiv.org/abs/2408.16116} {arXiv:2408.16116 [gr-qc]} \BibitemShut {NoStop}%
\bibitem [{\citenamefont {Matyjasek}(2021)}]{Matyjasek:2021xfg}%
  \BibitemOpen
  \bibfield  {author} {\bibinfo {author} {\bibfnamefont {J.}~\bibnamefont {Matyjasek}},\ }\href {\doibase 10.1103/PhysRevD.104.084066} {\bibfield  {journal} {\bibinfo  {journal} {Phys. Rev. D}\ }\textbf {\bibinfo {volume} {104}},\ \bibinfo {pages} {084066} (\bibinfo {year} {2021})},\ \Eprint {http://arxiv.org/abs/2107.04815} {arXiv:2107.04815 [gr-qc]} \BibitemShut {NoStop}%
\bibitem [{\citenamefont {Konoplya}\ and\ \citenamefont {Abdalla}(2005)}]{Konoplya:2005sy}%
  \BibitemOpen
  \bibfield  {author} {\bibinfo {author} {\bibfnamefont {R.~A.}\ \bibnamefont {Konoplya}}\ and\ \bibinfo {author} {\bibfnamefont {E.}~\bibnamefont {Abdalla}},\ }\href {\doibase 10.1103/PhysRevD.71.084015} {\bibfield  {journal} {\bibinfo  {journal} {Phys. Rev. D}\ }\textbf {\bibinfo {volume} {71}},\ \bibinfo {pages} {084015} (\bibinfo {year} {2005})},\ \Eprint {http://arxiv.org/abs/hep-th/0503029} {arXiv:hep-th/0503029} \BibitemShut {NoStop}%
\bibitem [{\citenamefont {Skvortsova}(2024)}]{Skvortsova:2024wly}%
  \BibitemOpen
  \bibfield  {author} {\bibinfo {author} {\bibfnamefont {M.}~\bibnamefont {Skvortsova}},\ }\href {\doibase 10.1134/S020228932470018X} {\bibfield  {journal} {\bibinfo  {journal} {Grav. Cosmol.}\ }\textbf {\bibinfo {volume} {30}},\ \bibinfo {pages} {279} (\bibinfo {year} {2024})},\ \Eprint {http://arxiv.org/abs/2405.15807} {arXiv:2405.15807 [gr-qc]} \BibitemShut {NoStop}%
\bibitem [{\citenamefont {Dubinsky}(2024{\natexlab{b}})}]{Dubinsky:2024aeu}%
  \BibitemOpen
  \bibfield  {author} {\bibinfo {author} {\bibfnamefont {A.}~\bibnamefont {Dubinsky}},\ }\href {\doibase 10.1016/j.dark.2024.101657} {\bibfield  {journal} {\bibinfo  {journal} {Phys. Dark Univ.}\ }\textbf {\bibinfo {volume} {46}},\ \bibinfo {pages} {101657} (\bibinfo {year} {2024}{\natexlab{b}})},\ \Eprint {http://arxiv.org/abs/2405.08262} {arXiv:2405.08262 [gr-qc]} \BibitemShut {NoStop}%
\bibitem [{\citenamefont {Dubinsky}(2024{\natexlab{c}})}]{Dubinsky:2024fvi}%
  \BibitemOpen
  \bibfield  {author} {\bibinfo {author} {\bibfnamefont {A.}~\bibnamefont {Dubinsky}},\ }\href {\doibase 10.13140/RG.2.2.35132.24961} {\  (\bibinfo {year} {2024}{\natexlab{c}}),\ 10.13140/RG.2.2.35132.24961},\ \Eprint {http://arxiv.org/abs/2405.13552} {arXiv:2405.13552 [gr-qc]} \BibitemShut {NoStop}%
\bibitem [{\citenamefont {Konoplya}(2002)}]{Konoplya:2001ji}%
  \BibitemOpen
  \bibfield  {author} {\bibinfo {author} {\bibfnamefont {R.~A.}\ \bibnamefont {Konoplya}},\ }\href {\doibase 10.1023/A:1015347628961} {\bibfield  {journal} {\bibinfo  {journal} {Gen. Rel. Grav.}\ }\textbf {\bibinfo {volume} {34}},\ \bibinfo {pages} {329} (\bibinfo {year} {2002})},\ \Eprint {http://arxiv.org/abs/gr-qc/0109096} {arXiv:gr-qc/0109096} \BibitemShut {NoStop}%
\bibitem [{\citenamefont {Konoplya}\ and\ \citenamefont {Zhidenko}(2007)}]{Konoplya:2006ar}%
  \BibitemOpen
  \bibfield  {author} {\bibinfo {author} {\bibfnamefont {R.~A.}\ \bibnamefont {Konoplya}}\ and\ \bibinfo {author} {\bibfnamefont {A.}~\bibnamefont {Zhidenko}},\ }\href {\doibase 10.1016/j.physletb.2007.03.018} {\bibfield  {journal} {\bibinfo  {journal} {Phys. Lett. B}\ }\textbf {\bibinfo {volume} {648}},\ \bibinfo {pages} {236} (\bibinfo {year} {2007})},\ \Eprint {http://arxiv.org/abs/hep-th/0611226} {arXiv:hep-th/0611226} \BibitemShut {NoStop}%
\bibitem [{\citenamefont {Bolokhov}(2024{\natexlab{b}})}]{Bolokhov:2023bwm}%
  \BibitemOpen
  \bibfield  {author} {\bibinfo {author} {\bibfnamefont {S.~V.}\ \bibnamefont {Bolokhov}},\ }\href {\doibase 10.1103/PhysRevD.110.024010} {\bibfield  {journal} {\bibinfo  {journal} {Phys. Rev. D}\ }\textbf {\bibinfo {volume} {110}},\ \bibinfo {pages} {024010} (\bibinfo {year} {2024}{\natexlab{b}})},\ \Eprint {http://arxiv.org/abs/2311.05503} {arXiv:2311.05503 [gr-qc]} \BibitemShut {NoStop}%
\bibitem [{\citenamefont {Kodama}\ \emph {et~al.}(2010)\citenamefont {Kodama}, \citenamefont {Konoplya},\ and\ \citenamefont {Zhidenko}}]{Kodama:2009bf}%
  \BibitemOpen
  \bibfield  {author} {\bibinfo {author} {\bibfnamefont {H.}~\bibnamefont {Kodama}}, \bibinfo {author} {\bibfnamefont {R.~A.}\ \bibnamefont {Konoplya}}, \ and\ \bibinfo {author} {\bibfnamefont {A.}~\bibnamefont {Zhidenko}},\ }\href {\doibase 10.1103/PhysRevD.81.044007} {\bibfield  {journal} {\bibinfo  {journal} {Phys. Rev. D}\ }\textbf {\bibinfo {volume} {81}},\ \bibinfo {pages} {044007} (\bibinfo {year} {2010})},\ \Eprint {http://arxiv.org/abs/0904.2154} {arXiv:0904.2154 [gr-qc]} \BibitemShut {NoStop}%
\bibitem [{\citenamefont {Konoplya}\ \emph {et~al.}(2019)\citenamefont {Konoplya}, \citenamefont {Zhidenko},\ and\ \citenamefont {Zinhailo}}]{Konoplya:2019hlu}%
  \BibitemOpen
  \bibfield  {author} {\bibinfo {author} {\bibfnamefont {R.~A.}\ \bibnamefont {Konoplya}}, \bibinfo {author} {\bibfnamefont {A.}~\bibnamefont {Zhidenko}}, \ and\ \bibinfo {author} {\bibfnamefont {A.~F.}\ \bibnamefont {Zinhailo}},\ }\href {\doibase 10.1088/1361-6382/ab2e25} {\bibfield  {journal} {\bibinfo  {journal} {Class. Quant. Grav.}\ }\textbf {\bibinfo {volume} {36}},\ \bibinfo {pages} {155002} (\bibinfo {year} {2019})},\ \Eprint {http://arxiv.org/abs/1904.10333} {arXiv:1904.10333 [gr-qc]} \BibitemShut {NoStop}%
\bibitem [{\citenamefont {Iyer}\ and\ \citenamefont {Will}(1987)}]{Iyer:1986np}%
  \BibitemOpen
  \bibfield  {author} {\bibinfo {author} {\bibfnamefont {S.}~\bibnamefont {Iyer}}\ and\ \bibinfo {author} {\bibfnamefont {C.~M.}\ \bibnamefont {Will}},\ }\href {\doibase 10.1103/PhysRevD.35.3621} {\bibfield  {journal} {\bibinfo  {journal} {Phys. Rev. D}\ }\textbf {\bibinfo {volume} {35}},\ \bibinfo {pages} {3621} (\bibinfo {year} {1987})}\BibitemShut {NoStop}%
\bibitem [{\citenamefont {Konoplya}(2003)}]{Konoplya:2003ii}%
  \BibitemOpen
  \bibfield  {author} {\bibinfo {author} {\bibfnamefont {R.~A.}\ \bibnamefont {Konoplya}},\ }\href {\doibase 10.1103/PhysRevD.68.024018} {\bibfield  {journal} {\bibinfo  {journal} {Phys. Rev. D}\ }\textbf {\bibinfo {volume} {68}},\ \bibinfo {pages} {024018} (\bibinfo {year} {2003})},\ \Eprint {http://arxiv.org/abs/gr-qc/0303052} {arXiv:gr-qc/0303052} \BibitemShut {NoStop}%
\bibitem [{\citenamefont {Matyjasek}\ and\ \citenamefont {Opala}(2017)}]{Matyjasek:2017psv}%
  \BibitemOpen
  \bibfield  {author} {\bibinfo {author} {\bibfnamefont {J.}~\bibnamefont {Matyjasek}}\ and\ \bibinfo {author} {\bibfnamefont {M.}~\bibnamefont {Opala}},\ }\href {\doibase 10.1103/PhysRevD.96.024011} {\bibfield  {journal} {\bibinfo  {journal} {Phys. Rev. D}\ }\textbf {\bibinfo {volume} {96}},\ \bibinfo {pages} {024011} (\bibinfo {year} {2017})},\ \Eprint {http://arxiv.org/abs/1704.00361} {arXiv:1704.00361 [gr-qc]} \BibitemShut {NoStop}%
\bibitem [{\citenamefont {Cardoso}\ \emph {et~al.}(2009)\citenamefont {Cardoso}, \citenamefont {Miranda}, \citenamefont {Berti}, \citenamefont {Witek},\ and\ \citenamefont {Zanchin}}]{Cardoso:2008bp}%
  \BibitemOpen
  \bibfield  {author} {\bibinfo {author} {\bibfnamefont {V.}~\bibnamefont {Cardoso}}, \bibinfo {author} {\bibfnamefont {A.~S.}\ \bibnamefont {Miranda}}, \bibinfo {author} {\bibfnamefont {E.}~\bibnamefont {Berti}}, \bibinfo {author} {\bibfnamefont {H.}~\bibnamefont {Witek}}, \ and\ \bibinfo {author} {\bibfnamefont {V.~T.}\ \bibnamefont {Zanchin}},\ }\href {\doibase 10.1103/PhysRevD.79.064016} {\bibfield  {journal} {\bibinfo  {journal} {Phys. Rev. D}\ }\textbf {\bibinfo {volume} {79}},\ \bibinfo {pages} {064016} (\bibinfo {year} {2009})},\ \Eprint {http://arxiv.org/abs/0812.1806} {arXiv:0812.1806 [hep-th]} \BibitemShut {NoStop}%
\bibitem [{\citenamefont {Konoplya}\ and\ \citenamefont {Stuchlík}(2017)}]{Konoplya:2017wot}%
  \BibitemOpen
  \bibfield  {author} {\bibinfo {author} {\bibfnamefont {R.~A.}\ \bibnamefont {Konoplya}}\ and\ \bibinfo {author} {\bibfnamefont {Z.}~\bibnamefont {Stuchlík}},\ }\href {\doibase 10.1016/j.physletb.2017.06.015} {\bibfield  {journal} {\bibinfo  {journal} {Phys. Lett. B}\ }\textbf {\bibinfo {volume} {771}},\ \bibinfo {pages} {597} (\bibinfo {year} {2017})},\ \Eprint {http://arxiv.org/abs/1705.05928} {arXiv:1705.05928 [gr-qc]} \BibitemShut {NoStop}%
\bibitem [{\citenamefont {Konoplya}(2023)}]{Konoplya:2022gjp}%
  \BibitemOpen
  \bibfield  {author} {\bibinfo {author} {\bibfnamefont {R.~A.}\ \bibnamefont {Konoplya}},\ }\href {\doibase 10.1016/j.physletb.2023.137674} {\bibfield  {journal} {\bibinfo  {journal} {Phys. Lett. B}\ }\textbf {\bibinfo {volume} {838}},\ \bibinfo {pages} {137674} (\bibinfo {year} {2023})},\ \Eprint {http://arxiv.org/abs/2210.08373} {arXiv:2210.08373 [gr-qc]} \BibitemShut {NoStop}%
\bibitem [{\citenamefont {Bolokhov}(2024{\natexlab{c}})}]{Bolokhov:2023dxq}%
  \BibitemOpen
  \bibfield  {author} {\bibinfo {author} {\bibfnamefont {S.~V.}\ \bibnamefont {Bolokhov}},\ }\href {\doibase 10.1016/j.physletb.2024.138879} {\bibfield  {journal} {\bibinfo  {journal} {Phys. Lett. B}\ }\textbf {\bibinfo {volume} {856}},\ \bibinfo {pages} {138879} (\bibinfo {year} {2024}{\natexlab{c}})},\ \Eprint {http://arxiv.org/abs/2310.12326} {arXiv:2310.12326 [gr-qc]} \BibitemShut {NoStop}%
\bibitem [{\citenamefont {Gleiser}\ and\ \citenamefont {Dotti}(2005)}]{Gleiser:2005ra}%
  \BibitemOpen
  \bibfield  {author} {\bibinfo {author} {\bibfnamefont {R.~J.}\ \bibnamefont {Gleiser}}\ and\ \bibinfo {author} {\bibfnamefont {G.}~\bibnamefont {Dotti}},\ }\href {\doibase 10.1103/PhysRevD.72.124002} {\bibfield  {journal} {\bibinfo  {journal} {Phys. Rev. D}\ }\textbf {\bibinfo {volume} {72}},\ \bibinfo {pages} {124002} (\bibinfo {year} {2005})},\ \Eprint {http://arxiv.org/abs/gr-qc/0510069} {arXiv:gr-qc/0510069} \BibitemShut {NoStop}%
\bibitem [{\citenamefont {Takahashi}\ and\ \citenamefont {Soda}(2012)}]{Takahashi:2011du}%
  \BibitemOpen
  \bibfield  {author} {\bibinfo {author} {\bibfnamefont {T.}~\bibnamefont {Takahashi}}\ and\ \bibinfo {author} {\bibfnamefont {J.}~\bibnamefont {Soda}},\ }\href {\doibase 10.1088/0264-9381/29/3/035008} {\bibfield  {journal} {\bibinfo  {journal} {Class. Quant. Grav.}\ }\textbf {\bibinfo {volume} {29}},\ \bibinfo {pages} {035008} (\bibinfo {year} {2012})},\ \Eprint {http://arxiv.org/abs/1108.5041} {arXiv:1108.5041 [hep-th]} \BibitemShut {NoStop}%
\bibitem [{\citenamefont {Konoplya}\ and\ \citenamefont {Zhidenko}(2017)}]{Konoplya:2017lhs}%
  \BibitemOpen
  \bibfield  {author} {\bibinfo {author} {\bibfnamefont {R.~A.}\ \bibnamefont {Konoplya}}\ and\ \bibinfo {author} {\bibfnamefont {A.}~\bibnamefont {Zhidenko}},\ }\href {\doibase 10.1088/1475-7516/2017/05/050} {\bibfield  {journal} {\bibinfo  {journal} {JCAP}\ }\textbf {\bibinfo {volume} {05}},\ \bibinfo {pages} {050} (\bibinfo {year} {2017})},\ \Eprint {http://arxiv.org/abs/1705.01656} {arXiv:1705.01656 [hep-th]} \BibitemShut {NoStop}%
\bibitem [{\citenamefont {Davey}\ \emph {et~al.}(2023)\citenamefont {Davey}, \citenamefont {Dias},\ and\ \citenamefont {Santos}}]{Davey:2023fin}%
  \BibitemOpen
  \bibfield  {author} {\bibinfo {author} {\bibfnamefont {A.}~\bibnamefont {Davey}}, \bibinfo {author} {\bibfnamefont {O.~J.~C.}\ \bibnamefont {Dias}}, \ and\ \bibinfo {author} {\bibfnamefont {J.~E.}\ \bibnamefont {Santos}},\ }\href {\doibase 10.1007/JHEP12(2023)101} {\bibfield  {journal} {\bibinfo  {journal} {JHEP}\ }\textbf {\bibinfo {volume} {12}},\ \bibinfo {pages} {101} (\bibinfo {year} {2023})},\ \Eprint {http://arxiv.org/abs/2305.11216} {arXiv:2305.11216 [gr-qc]} \BibitemShut {NoStop}%
\end{thebibliography}%
\end{document}